\documentclass{Pos}
\usepackage{epsf}
\usepackage{amsmath}

\def\beq{\begin{equation}}
\def\eeq{\end{equation}}
\def\bea{\begin{eqnarray}}
\def\eea{\end{eqnarray}}
\def\beqa{\begin{equation}\begin{array}{l}}
\def\eeqa{\end{array}\end{equation}}
\def\eqlab#1{\label{eq:#1}}
\def\figlab#1{\label{fig:#1}}

\def\eref#1{(\ref{eq:#1})}
\def\Eqref#1{Eq.~(\ref{eq:#1})}

\def\Figref#1{Fig.~\ref{fig:#1}}

\def\half{\mbox{\small{$\frac{1}{2}$}}}

\def\quarter{\mbox{\small{$\frac{1}{4}$}}}

\def\al{\alpha}
\def\be{\beta}
\def\ga{\gamma} 
\def\de{\delta} \def\De{\Delta}
  \def\eps{\epsilon}

\def\la{\lambda} \def\La{{\Lambda}}

\def\si{\sigma} 
\def\th{\theta}

\def\pa{\partial}

\def\pa{\partial}

\def\nn{\nonumber}

\title{Chiral EFT with a resonance and heavy fields }

\ShortTitle{Chiral EFT with a resonance and heavy fields }

\author{\speaker{Vladimir Pascalutsa}
       \\
       Institut f\"ur Kernphysik, Johannes Gutenberg Universit\"at, Mainz D-55099, Germany\\
       }

\abstract{Several conceptual points concerning the inclusion of the $\Delta$(1232) resonance in the framework of chiral effective-field  theory are discussed, with an emphasis on
the problem of power counting in the baryon sector in general. I also formulate
a new dispersion relation in the pion-mass squared (or, the quark mass) and make a
link between the power counting and the analytic properties of chiral expansion. A controversy regarding the determination of the
proton's magnetic polarizability from Compton-scattering data is stressed here as well.}

\FullConference{6th International Workshop on Chiral Dynamics\\
		 July 6-10 2009\\
		 Bern, Switzerland}

\begin{document}

\section{Introduction}
The limit of applicability of chiral perturbation theory 
($\chi$PT)~\cite{Weinberg:1978kz,Gasser:1983yg,GSS89} is ultimately set
by the scale of spontaneous chiral symmetry breaking: 
$\Lambda_{\chi SB} \simeq 4 \pi f_\pi\sim 1$ GeV, but not only. Resonances
with excitation energy lower than 1 GeV, e.g.\ $\rho$(770) or $\Delta$(1232),
set a lower limit, if not included explicitly. The $\Delta$(1232)
is especially important because of its very low excitation energy, as defined by
the $\Delta$-nucleon mass splitting:
\begin{equation}
\varDelta \equiv M_\Delta - M_N \approx (1232-939) \,\mbox{MeV} = 293 \,\mbox{MeV.}
\end{equation}
This means we can expect an early breakdown of $\chi$PT in the baryon sector,
on one hand, but an easy fit of the $\Delta$ into the $\chi$PT power-counting
scheme ($\varDelta \ll \Lambda_{\chi SB}$), on the other.

A first work on the inclusion of $\Delta$-resonance and, more generally, the
decuplet fields in $\chi$PT was done by Jenkins and Manohar \cite{Jenkins:1991es}, 
who at the same time developed the ``heavy-baryon" (HB) expansion~\cite{JeM91a}. 
They counted the $\Delta$-excitation scale to be of the same order as other
light scales in the theory, i.e., Goldstone-boson momenta and masses.
For the two-flavor QCD, this hierarchy of scales, 
\begin{equation}
\varDelta \sim p \sim m_\pi \ll \Lambda_{\chi SB}\,,
\end{equation}
 results in the ``small scale expansion" (SSE) \cite{Hemmert:1996xg}.

Alternatively, one can count the resonance excitation scale to be different from
the pion mass, i.e.,
 \begin{equation}
 m_\pi < \varDelta \ll \Lambda_{\chi SB}\, .
\end{equation}
This is an example of effective-field theory (EFT) with two distinct light scales.
The power counting of graphs will then depend on
whether the typical momenta are comparable to $m_\pi$ or to $\varDelta$.
The expansion can be carried out in terms of one
small parameter, e.g.,
 \begin{equation}
 \delta = \frac{\varDelta}{  \Lambda_{\chi SB}}\ll 1\, .
\end{equation}
Then, $m_\pi / \Lambda_{\chi SB}$ should count as $\delta$ to some power greater than one.
The simplest is to take an integer power:
\beq
\frac{m_\pi}{  \Lambda_{\chi SB}} \sim \delta^2 .
\eeq
This counting scheme goes under the name of ``$\delta$-expansion" \cite{Pascalutsa:2003zk}.

The main advantage of the $\delta$-expansion over the SSE is that it provides a more
adequate counting of the resonant contributions and
a power-counting argument to sum a subset of graphs generating the resonance width.

In Sect.~4 we shall see a brief account of one recent applications of the $\delta$-expansion,
a new calculation of the $\Delta$-resonance effect on
the nucleon polarizabilities and Compton scattering off protons \cite{Lensky:2008re}.
More applications can be found elsewhere \cite{Pascalutsa:2007wb,Geng:2008bm,Geng:2009hh,Long:2009wq}, including these proceedings~\cite{McGovern:2009sw}.
The other purpose of this paper is to remark on
two, quite unrelated, consistency problems of $\chi$PT in the baryon sector (B$\chi$PT). 
One concerns the treatment of  higher-spin fields  (Sect.~2), and the other is about
the power counting (Sect.~3 and 4).

\section{Higher-spin  fields}
Including the baryon field in the chiral Lagrangian, one sooner or later faces
the consistency problems of higher-spin field theory.
The $\Delta$(1232), being a spin-3/2 state, can be represented
by a Rarita-Schwinger (RS) vector-spinor, $\psi_\alpha(x)$, with the following
free Lagrangian:
\beq
\mathcal{L}_{RS}= \bar \psi_\alpha \, ( i\, \gamma^{\alpha\beta\varrho} \partial_\varrho - M_\Delta \,\gamma^{\alpha\beta}) \, \psi_\beta,
\eeq
where $\gamma^{\alpha\beta\rho} $ and $\ga^{\al\be}$ are 
totally antisymmetrized products of three and two Dirac matrices.
The Lagrangian consists of a kinetic term, which is invariant under
a gauge symmetry:
\beq
\psi_\al \to \psi_\al + \pa_\al \eps
\eeq
(with a spinor $\eps$), and a mass term, which breaks
the gauge symmetry. 
This formalism provides a proper field-theoretic description of a spin-3/2
particle. The symmetry ensures that  the massless particle has 2 spin degrees of freedom,
while the mass term breaks it such as to raise the number of spin
degrees of freedom to 4. 
This pattern has to be preserved by interactions
of this field, but such a consistency criterion proved to be tough 
to fulfill. 

For instance, the usual minimal substitution 
of the electromagnetic field, $\pa_\rho \to \pa_\rho + i e A_\rho$, leads
to U(1)-invariant theory, but at expense of loss of the spin-3/2 gauge symmetry
of massless theory. As the result, all hell,
with its negative-norm states \cite{Johnson:1961vt}, 
superluminal modes \cite{Velo:1969bt}, etc.~\cite{Deser:2000dz}, breaks loose.
Naive attempts to restore the spin-3/2 gauge symmetry break the U(1)
gauge symmetry, and so on.
In fact, there are `no-go theorems'  forbidding a consistent
coupling of a spin-3/2 field to electromagnetism
without gravity, see e.g., \cite{Weinberg:1980kq}. 

This situation is frustrating, especially since  we would like to couple
the  $\Delta$'s to pions too, and so, chiral symmetry is one more symmetry to worry about. 
Fortunately, `locality' is one of the principles that underlines
the `no-go theorem', and, given that the EFT framework
is essentially non-local, we have a way to work around it.
One method has been outlined in Ref.~\cite{Pascalutsa:2006up}
(Sect.~4.2 therein), and a similar method  has been developed in
parallel~\cite{Krebs:2008zb}. However, a complete closed-form solution to this problem
is still lacking.

\section{Heavy fields and dispersion in the pion mass}
Another important issue of concern is
the treatment of heavy fields in $\chi$PT. This
problem comes already with the inclusion of the nucleon field. 
A key question is: "how to count derivatives of the nucleon field?"
The nucleon is heavy ($M_{\,N} \sim \La_{\chi SB}$), and hence
the time (0th) component of the nucleon derivative, or momentum,
is much greater than the spatial components:
\beq
\pa_i N(x) \ll \pa_0 N(x),
\eeq 
or, in the momentum space, $p \ll \sqrt{M_N^2 +p^2}$, for an on-shell nucleon.
It would be correct to count the 0th component as $\mathcal{O}(1)$, while
the spatial components as $\mathcal{O}(p)$, but this counting obviously 
does not respect the Lorentz invariance. 

In a Lorentz-invariant formulation, $\pa_\mu N$ counts 
as $\mathcal{O}(1)$, except when in a particular combination, $(i \pa\!\!\!\!\!/ - M_N)N$,
which counts as $\mathcal{O}(p)$. This counting has a consistency problem,
as can be seen from the following example. Consider an expression,
$ P_\mu - M_N \ga_\mu$, 
where $P_\mu$ here is the nucleon four-momentum which, as $\ga_\mu$ and $M_N$, 
counts as 1. The counting of this expression,  as a whole, will unfortunately depend on
how it is contracted. E.g., whether contracted with $p$ or $\gamma$ we have:
\bea
P^\mu (P_\mu - M_N \ga_\mu) = P \!\!\!\!\!/ \, ( P \!\!\!\!\!/ \, - M_N) & \sim & \mathcal{O}(p), \nn\\
\ga^\mu (P_\mu - M_N \ga_\mu) = -3 M_N + ( P \!\!\!\!\!/ \, - M_N) &\sim & \mathcal{O}(1). \nn
\eea
This inconsistency leads eventually to the appearance of nominally
lower-order or higher-order contributions than ones expected from power-counting
\cite{GSS89}.

The heavy-baryon (HB) expansion of Jenkins and Manohar \cite{JeM91a} overcomes this problem, 
but again, at the expense of manifest Lorentz-invariance. In HB$\chi$PT one writes
\beq
P_\mu = M_N v_\mu + \ell_\mu 
\eeq 
with $v=(1, 0, 0, 0)$, which allows to assign a consistent power to $\ell$.

More recently it is becoming increasingly clear that the power-counting
problem of Lorentz-invariant formulation is not very severe \cite{Becher:1999he}, or perhaps not a problem at all  \cite{Gegelia:1999gf}.
The lower-order `power-counting violating' contributions 
come out to be analytic in quark masses, 
and therefore match the contributions that come multiplying the low-energy constants
(LECs), and as result, 
do not play any role other than renormalizing the LECs.
The higher-order contributions, on the other hand, can be both analytic and non-analytic 
in quark masses.
Their analytic parts may contain ultra-violet divergencies, so one needs to define
the renormalization scheme for the higher-order LECs, before they
actually appear in the calculation. The non-analytic parts are most interesting,
as they may come with unnaturally large coefficients, and therefore cannot be dismissed
as `higher order' at all. 

\begin{figure}[bt]
\begin{minipage}[c]{.32\linewidth}
\centerline{  \epsfxsize=4cm
  \epsffile{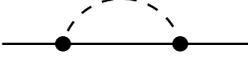} 
}
\end{minipage}
\hspace{.06\linewidth}%
\begin{minipage}[c]{.57\linewidth}
\caption{The nucleon self-energy contribution at order $p^3$.}
\figlab{Nselfen}
\end{minipage}
\end{figure}

This discussion is nicely illustrated by the classic example of chiral corrections to 
the nucleon mass. Up to $\mathcal{O}(p^3)$  this expansion is given by
\beq
\eqlab{expansion}
M_N = {M}_{N0} - 4 c_1 m_\pi^2 + \Sigma^{(3)}_N,
\eeq
where ${M}_{N0}$ and $c_1$ are LECs which, supposedly,  represent the values
of nucleon mass and $\pi N$ $\si$-term 
in the chiral limit. The last term is the (leading) 3rd-order self-energy correction, 
\Figref{Nselfen}:
\begin{subequations}
\bea
\Sigma^{(3)}_N & = &\left. i \,\frac{3 g_A^2}{4f_\pi^2}  \int \!\frac{d^4 k }{(2\pi)^4}
\frac{k \!\!\!\!/ \gamma_5 ( p\!\!\!\!/ - k\!\!\!\!/+M_{N}) k \!\!\!\!/ \gamma_5}{(k^2-m_\pi^2) [(p-k)^2-M_{N}^2]} \right|_{p\!\!\!\!/ = M_N}\\
& \stackrel{\mathrm{dim reg}}{=} & 
\frac{3 g_A^2}{4f_\pi^2} \frac{M_{N}^3}{(4\pi)^2} \int\nolimits_0^1\! dx\, \Big\{
 [x^2+\mu^2 (1-x)] \left(L_\eps+\ln [x^2+\mu^2 (1-x)]\right) \nn\\
& & \hskip2.5cm + \, [2x^2-\mu^2 (2+x)]  \left(L_\eps+1+\ln [x^2+\mu^2(1-x) ]\right)-3
L_\eps\Big\},
\eqlab{selfen}
\eea
\end{subequations}
where $\mu = m_\pi/M_{N}$, while $L_\eps = -1/\eps -1+ \gamma_E - \ln(4\pi \La/M_{N})$
exhibits the ultraviolet divergence as $\eps=(4-d)/2 \to 0$, with $d$ being the number of dimensions, $\La$  the scale of dimensional regularization, and $\gamma_{E}$ the Euler's constant. Note that we took the physical nucleon mass for the on-mass-shell condition, as well
as for the propagator pole, and not the chiral-limit mass $M_{N0}$, 
which comes from the Lagrangian.
There are several reasons for that  (for one, $M_N$ is the ``known known" here), but
in any case the difference between doing it one way or the other is
of higher order.

After the integration over $x$ we obtain: 
\begin{subequations}
\bea
&& \Sigma^{(3)}_N = \frac{3 g_A^2 M_{N}^3}{2(4\pi f_\pi)^2}\left\{- L_\eps
+\left(1-L_\eps\right)\mu^2 \right\}\,+\,\overline \Sigma^{(3)}_N,
\eqlab{low} \\
\mbox{with} && \overline \Sigma^{(3)}_N = - \frac{3 g_A^2 M_{N}^3}{(4\pi f_\pi)^2}
 \Big( \mu^3 \sqrt{1-\quarter \mu^2} \, \,\arccos \half \mu + \quarter \mu^4\,  \ln \mu^2 \Big)\nn\\
 && \,\,\,\, = \, - \frac{3 g_A^2}{(4\pi f_\pi)^2} \frac{1}{2}\Big[ \pi \, m_\pi^3 - (m_\pi^4/M_N)
  (1- \ln m_\pi/M_N ) + \mathcal{O}(m_\pi^5) \Big] .
  \eqlab{renorm}
\eea
\end{subequations}
Now we can see the problem explicitly. While the power-counting 
of the graph (\Figref{Nselfen}) gives order 3, the result contains
both lower and higher powers of the light scale, $m_{\,\pi}$. 

The higher-order terms should not be a problem. Formally  we can either keep them or not
without an effect to the accuracy with which we work. There are cases where 
it is not as simple as that. One such case is considered in the next section. 

The lower-order terms,
written out in \Eqref{low}, have been of a bigger concern \cite{GSS89}.
Fortunately, they are of the same form as  the first two terms
 in the expansion of nucleon mass,  \Eqref{expansion}. Chiral symmetry
ensures this ``miracle" happens every time. The troublesome lower-order
terms can thus be absorbed into a renormalization of the available LECs ---
a view introduced by Gegelia and Japaridze \cite{Gegelia:1999gf}. 
In fact,
these terms {\it must} be absorbed, if $M_{N0}$ and $c_1$ are really to represent the values of nucleon mass and $\si$-term in the chiral limit. As a result, 
\beq
\eqlab{expansion}
M_N = {M}_{N0} - 4 c_1 m_\pi^2 + \overline\Sigma^{(3)}_N,
\eeq
and all is well,  from the power-counting point of view.
The only question left (in some expert's minds)
is whether
these LECs will be renormalized in exact same amounts in calculations of other
quantities at
this order. In my view, again, the symmetries ensure this is so. 
I am not aware of an example to the contrary.

Alternatively, the HB formalism \cite{JeM91a} yields right away the following
expression for the graph of \Figref{Nselfen}:
\beq
\Sigma^{(3)HB}_N = - \frac{3 g_A^2}{(4\pi f_\pi)^2} \frac{1}{2}\pi \, m_\pi^3, 
\eeq
i.e., only the first term in the expansion of the renormalized self-energy, \Eqref{renorm}.
So, no lower-order terms are present (in dimensional regularization!), 
no higher-order terms either:
a perfect consistency with power counting. However, as practice shows, in too many cases
the thus neglected higher-order (in $p/M_N$) terms are not that small. 
Unlike in the above-considered example of nucleon mass, the higher powers of $m_\pi/M_N$
can come with `unnaturally large' coefficients. In these cases,
the HB expansion demonstrates poor convergence. 
One such case --- the nucleon polarizabilities --- will be considered below, but first, I would
like to introduce a principle of {\it analyticity} of the chiral expansion.

For this purpose I would like to have a dispersion relation in the
variable $t=m_{\,\,\,\pi}^{\,\,\,2}$. It is clear that for negative $t$, the chiral-loop graphs of
the type in \Figref{Nselfen} will have an imaginary part, reflecting the possibility
of decay of the nucleon into itself and a tachionic pion,  and hence there is a cut
extending from $t=0$ to $t=-\infty$. In the rest of the complex $t$ plane, 
 we can expect an analytic dependence. A dispersion-relation for a quantity such as
nucleon self-energy must then read:
\beq
\mathrm{Re}\, \Sigma_N(t)  = -\frac{1}{\pi} \int\limits_{-\infty}^0 dt' \,\frac{\mathrm{Im}\, \Sigma_N (t') }{t'-t}
\eeq
In the above example of 3rd order self-energy, we can easily find the imaginary part from 
\Eqref{selfen}, if we restore the $i \eps$ prescription and use $\ln(-1+i\eps) = i\pi$,
\beq
\mathrm{Im}\, \Sigma_N^{(3)} (t) = 
 \frac{3 g_A^2}{(4\pi f_\pi)^2} \frac{\pi}{2} \left[ - (-t)^{3/2} \left( 1-\frac{t}{4M_N^2} \right)^{1/2}
\! + \,\frac{t^2}{2 M_N}\right] \, \th(-t)\,.
\eeq
According to the expansion \Eqref{expansion}, we should be making at least two subtractions
at $t=0$, and hence
\beq
\mathrm{Re}\, \overline \Sigma_N(t)  = \mathrm{Re}\, \Sigma_N(t)  - 
 \mathrm{Re}\, \Sigma_N(0) -  \mathrm{Re}\, \Sigma_N' (0)\, \, t \, =\, 
 -\frac{1}{\pi} \int\limits_{-\infty}^0 dt' \,\frac{\mathrm{Im}\, \Sigma_N (t') }{t'-t} \left(\frac{t}{t'}\right)^2.
\eeq
Substituting the expression for the imaginary part, and taking $t=m_{\,\,\,\pi}^{\,\,\,2}$, we
indeed recover the  result of \Eqref{renorm}, therefore validating the analyticity assumptions
on one hand, and  revealing the intricate nature of  the `higher-order terms' on the other.

\section{Compton scattering and proton polarizabilities}

\begin{figure}[tb]
\begin{minipage}[c]{.4\linewidth}
\centerline{\epsfclipon  \epsfxsize=6cm%
  \epsffile{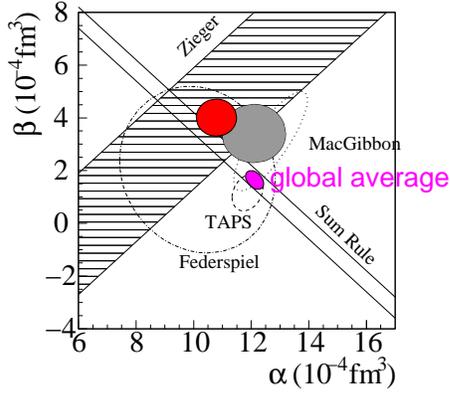} 
}
\end{minipage}
\hspace{.1\linewidth}%
\begin{minipage}[c]{.45\linewidth}
\caption{The scalar polarizabilities of the proton. 
The results of HB$\chi$PT~\cite{Beane:2004ra} and B$\chi$PT~\cite{Lensky:2008re} are
 shown respectively by the grey and red blob. Experimental
results are from Federspiel et~al.~\cite{Federspiel:1991yd},
Zieger et al.~\cite{Zieger:1992jq}, MacGibbon et al.~\cite{MacG95},
and TAPS~\cite{MAMI01}.
`Sum Rule' indicates the Baldin sum rule constraint on $\alpha+\beta$~\cite{Bab98}.
`Global average' represents the PDG summary~\cite{PDG2006}.} 
\figlab{potato}
\end{minipage}
\end{figure}

The main aim of low-energy Compton scattering experiments on protons
and light nuclei in recent years has been to detect the nucleon {\it polarizabilities} \cite{Holstein:1992xr}. 
For the
scalar electric $\al$ and magnetic $\be$ polarizabilities of the proton, the phenomenology
seems to be in a very good shape, see `global average' in \Figref{potato}. That's 
why it is intriguing to see that these values are not entirely in agreement with two
recent $\chi$PT calculations (cf.\ the grey  \cite{Beane:2004ra}
and the red \cite{Lensky:2008re} blob in  \Figref{potato}). Note that, the
$\chi$PT analyses are not in disagreement with the experimental data
for cross-sections, as \Figref{fixE} shows,  for example, in the case of Ref.~\cite{Lensky:2008re}.
The principal differences with phenomenology arise  apparently at the stage
of interpreting the effects of polarizabilities in Compton observables. It is
important to sort out this disagreement in a near future, perhaps with the help of a round 
of new experiments at MAMI and HIGS.

\begin{figure}[t]
\begin{minipage}[c]{.57\linewidth}
\centerline{ \epsfclipon  
\epsfxsize=8.5cm%
  \epsffile{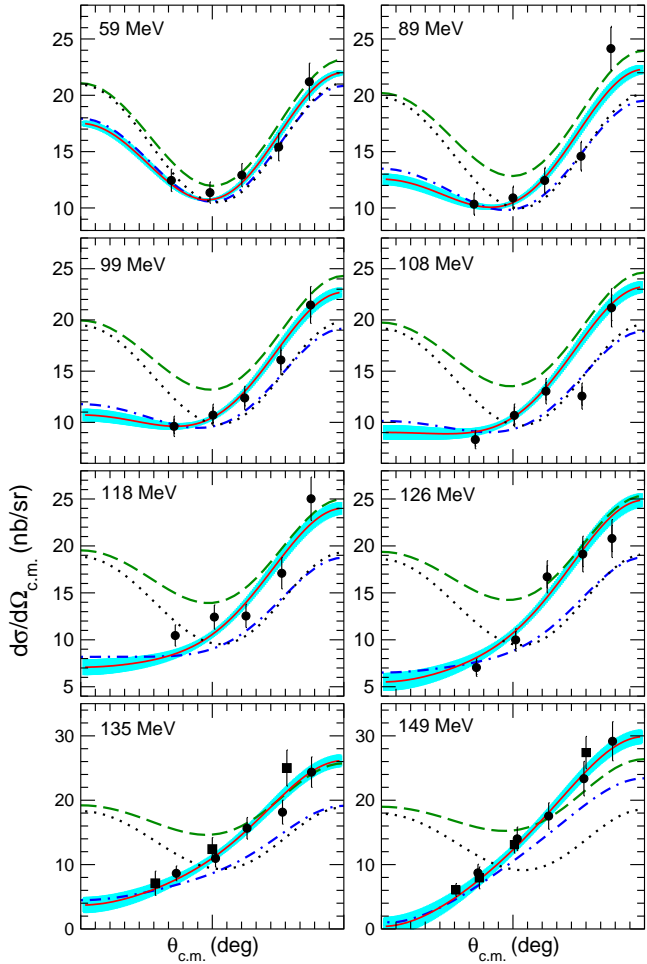} 
}
\end{minipage}
\hspace{.06\linewidth}%
\begin{minipage}[c]{.34\linewidth}
\caption{Angular dependence of 
the $\ga p\to \ga p$ differential cross-section in
the center-of-mass system for a fixed photon-beam energies
as specified for each panel. Data points are from SAL~\cite{Hal93} ---
filled squares, and MAMI~\cite{MAMI01} --- filled circles. The curves are:
Klein-Nishina --- dotted, Born graphs and WZW-anomaly --- green dashed,
adding the $p^3$ $\pi N$ loop contributions of B$\chi$PT
--- blue dash-dotted. The result of adding the $\Delta$
contributions, i.e., the complete NNLO result of Ref.~\cite{Lensky:2008re}, is shown by the red solid line with a band.}
\figlab{fixE}
\end{minipage}
\end{figure}

For now, however, I focus on the differences between the two $\chi$PT
calculations. The earlier one  \cite{Beane:2004ra} is done in HB$\chi$PT at order $p^4$.
The latest is a manifestly-covariant calculation at order $p^3$ and $p^4/\varDelta$,
hence includes the $\De$-isobar effects within the $\de$-counting scheme. Despite the similar results for polarizabilities,
the composition of these results order by order is quite different. In HB$\chi$PT
one obtains for the central values (in units of $10^{-4}$fm$^3$):
\begin{eqnarray}
 \alpha &=& \underbrace{12.2}_{\mathcal{O}(p^3)} + \underbrace{(-0.1)}_{\mathcal{O}(p^4)} = 12.1\,,\\
\beta&=&\underbrace{1.2}_{\mathcal{O}(p^3)} +\underbrace{2.2}_{\mathcal{O}(p^4)} =3.4  \,.
\end{eqnarray}
while in B$\chi$PT with $\Delta$'s:
\begin{eqnarray}
 \alpha &=& \underbrace{6.8}_{\mathcal{O}(p^3)} + \underbrace{(-0.1) + 4.1}_{\mathcal{O}(p^4/\varDelta)} = 10.8\,,\\
\beta&=&\underbrace{-1.8}_{\mathcal{O}(p^3)} +\underbrace{ 7.1-1.3}_{\mathcal{O}(p^4/\varDelta)} =4.0  \,.
\end{eqnarray}
The difference at the leading order comes precisely due to the `higher-order' terms.
For instance, for the magnetic polarizability at $\mathcal{O}(p^3)$, in one case we have:
\beq
\beta^{(3)HB}= \frac{e^2g_A^2}{768\pi^2 f_\pi^2 m_\pi }\,,
\eqlab{HBresult}
\eeq
while in the other:
\begin{eqnarray}
 \beta^{(3)}&=&\frac{e^2 g_A^2}{192\pi^3 f_\pi^2 M_N }
  \int\limits^1_0\! dx \, \bigg\{ 1 - \frac{(1-x)(1-3x)^2+x}{\mu^2 (1-x)+x^2} 
  -  \frac{x \mu^2 +x^2 [1-(1-x)(4-20x+21x^2)]}{\big[\mu^2 (1-x)+x^2\big]^2} 
\bigg\}\nn \\
&=&\frac{e^2g_A^2}{192\pi^3 f_\pi^2 M_N }\Big[\, \frac{\pi}{4\mu}+18\ln\mu+\frac{63}{2}
-\frac{981\pi}{32}\mu-\big(100\, \ln\mu+\frac{121}{6}\big)\mu^2+\ldots \Big].
\eqlab{expanded}
\end{eqnarray}
The first term in the expanded expression \eref{expanded}
is exactly the same as the HB result \eref{HBresult}, but how about the
higher-order terms. Their coefficients are at least a factor of 10 bigger than the
coefficient of the leading term. Given that the expansion parameter is $\mu \sim 1/7$,
there is simply no argument why these terms should be neglected.

As a consequence, the neat agreement of the HB $p^3$ result
with the empirical numbers for $\al$ and $\be$ should perhaps be viewed 
as a remarkable coincidence, rather than, as it's often viewed, 
a ``remarkable prediction" of HB$\chi$PT. 
In fact, the predictive power is what is first
of all compromised by the unnaturally large, higher order (in  HB-expansion) terms.
These terms will of course be recovered  in the higher-order HB calculations, but with
each order higher there will be an increasingly higher number of unknown LECs.
In contrast, the covariant results provide an example of how one gets to the important 
effects already in the lower-order calculations, before new LECs start to appear.

Now let's return to the dispersion relations in the pion mass squared.
Denoting $t=\mu^2$,
the dispersion relation for the magnetic polarizability should read
\beq
\mathrm{Re}\, \be(t)  = -\frac{1}{\pi} \int\limits_{-\infty}^0 dt' \,\frac{\mathrm{Im}\, \be (t') }{t'-t}\,.
\eqlab{DRbeta}
\eeq
 The imaginary
part at the 3rd order can be calculated from the first line of \Eqref{expanded}, 
\bea
\mathrm{Im}\, \be^{(3)} (t) &=& -\,
C\,\,
\mathrm{Im}\,  \int\limits^1_0\! dx\, \bigg\{ 
\frac{(1-x)(1-3x)^2 +x+\mbox{$\frac{x}{1-x}$} }{(1-x)t+x^2-i\eps} \nn\\
&&\hskip4cm - \frac{d}{dt } \frac{x \,t  +x^2 [1-(1-x)(4-20x+21x^2)]}{(1-x) [(1-x)t+x^2-i\eps]} 
\bigg\} \\
&=& - 
\frac{\pi C}{8\la^3} \big[ 2-72 \la +t(418\la-246) - t^2(316\la-471) +t^3(54\la-212) +27t^4\big],
\nn
\eea
where $C=\frac{e^2 g_A^2}{192\pi^3 f_\pi^2 M_N }$ and
$\la = \sqrt{-t\,(1-\quarter t)}$. At this order there are no counter-terms, hence no subtractions,
and indeed I have verified that the unsubtracted relation \eref{DRbeta} gives exactly the
same result as \Eqref{expanded}. 
The relation for the electric polarizability $\al$ has been verified in a similar fashion. These
 tests validate the analyticity assumption and elucidate the nature of the 'higher-order'  terms.
 
Finally, let me note that such dispersion relations,
 as well as the usual ones (in energy) \cite{Holstein:2005db}, do not hold in the framework of
  Infrared Regularization (IR)  \cite{Becher:1999he}.
  The IR loop-integrals will always, in addition to the unitarity cuts, have an unphysical
  cut. Although the unphysical cut lies far from the region of $\chi$PT applicability and
  therefore does not pose a threat to unitarity, it does make an impact, and as result,
  a set of the higher-order terms is altered. To me, this is a showstopper.  The only
  practical advantage of the manifest Lorentz-invariant formulation over the HB one
  is the account of `higher-order' terms which may, or may not, be unnaturally large. Giving up on analyticity, one has no principle to assess these terms reliably.

\section{Summary}

Here are some points which have been illustrated in this paper:
\begin{itemize}
\item The region of applicability of B$\chi$PT without the $\De$(1232)-baryon is: $p\ll 300$ MeV.
An explicit $\De(1232)$ is needed to extend this limit to substantially higher energies. Two
schemes are presently used to power-count the $\De$ contributions: SSE and $\de$-expansion.
\item Inclusion of heavy fields poses a difficulty with power counting in a Lorentz-invariant
formulation --- contributions of lower- and higher-order arise in a calculations given-order graph.
However, this is not a problem --- the lower-order contributions renormalize the available LECs,
while the higher-order ones are, in fact, required by analyticity and should be kept.
\item Dispersion relations in the pion-mass squared have been derived and are shown
to hold in the examples of lowest order chiral corrections to the nucleon mass and
polarizabiltities.
\item The present state-of-art $\chi$PT calculations of low-energy Compton scattering
are in a good agreement with experimental cross-sections,  but have an appreciable
discrepancy with PDG values for proton polarizabilities.
\end{itemize}


\begin{thebibliography}{99}
\bibitem{Weinberg:1978kz}
  S.~Weinberg,
  Physica A {\bf 96}, 327 (1979).

\bibitem{Gasser:1983yg}
  J.~Gasser and H.~Leutwyler,
  Annals Phys.\  {\bf 158} (1984) 142.
  
 \bibitem{GSS89}
J.~Gasser, M.~E.~Sainio and A.~Svarc,
Nucl.\ Phys.\ B {\bf 307}, 779 (1988).

\bibitem{Jenkins:1991es}
  E.~Jenkins and A.~V.~Manohar,
  Phys.\ Lett.\  B {\bf 259}, 353 (1991).

\bibitem{JeM91a}
E.~Jenkins and A.~V.~Manohar,
Phys.\ Lett.\ B {\bf 255}, 558 (1991).

\bibitem{Hemmert:1996xg}
  T.~R.~Hemmert, B.~R.~Holstein and J.~Kambor,
  Phys.\ Lett.\  B {\bf 395}, 89 (1997);
  J.\ Phys.\ G {\bf 24}, 1831 (1998).

   \bibitem{Pascalutsa:2003zk}
  V.~Pascalutsa and D.~R.~Phillips,
  Phys.\ Rev.\  C {\bf 67}, 055202 (2003).
  
\bibitem{Lensky:2008re}
  V.~Lensky and V.~Pascalutsa,
  Pisma Zh.\ Eksp.\ Teor.\ Fiz.\  {\bf 89}, 127 (2009)
  [JETP Lett.\  {\bf 89}, 108 (2009)];
 arXiv:0907.0451 [hep-ph], submitted to Eur. J. Phys. C.
  
\bibitem{Pascalutsa:2007wb}
  V.~Pascalutsa and M.~Vanderhaeghen,
    Phys.\ Rev.\ Lett.\  {\bf 95}, 232001 (2005);
 Phys.\ Rev.\  D {\bf 73}, 034003 (2006); 
   Phys.\ Rev.\ Lett.\  {\bf 94}, 102003 (2005);
 Phys.\ Rev.\  D {\bf 77}, 014027 (2008);
    Phys.\ Lett.\  B {\bf 636}, 31 (2006).

  \bibitem{Geng:2008bm}
  L.~S.~Geng, J.~Martin Camalich, L.~Alvarez-Ruso and M.~J.~Vicente Vacas,
  Phys.\ Rev.\  D {\bf 78}, 014011 (2008).

\bibitem{Geng:2009hh}
  L.~S.~Geng, J.~Martin Camalich and M.~J.~Vicente Vacas,
  Phys.\ Lett.\  B {\bf 676}, 63 (2009).

\bibitem{Long:2009wq}
  B.~Long and U.~van Kolck,
  arXiv:0907.4569 [hep-ph].

\bibitem{McGovern:2009sw}
  J.~A.~McGovern, H.~W.~Griesshammer, D.~R.~Phillips and D.~Shukla,
  arXiv:0910.1184 [nucl-th].

  
\bibitem{Johnson:1961vt}
K.~Johnson and E.~C.~Sudarshan,
Annals Phys.\  {\bf 13}, 126 (1961).
\bibitem{Velo:1969bt}
G.~Velo and D.~Zwanziger,
Phys.\ Rev.\  {\bf 186}, 1337 (1969).

\bibitem{Deser:2000dz}
  S.~Deser, V.~Pascalutsa and A.~Waldron,
  Phys.\ Rev.\  D {\bf 62}, 105031 (2000).


\bibitem{Weinberg:1980kq}
  S.~Weinberg and E.~Witten,
  Phys.\ Lett.\  B {\bf 96}, 59 (1980).
  
\bibitem{Pascalutsa:2006up}
  V.~Pascalutsa, M.~Vanderhaeghen and S.~N.~Yang,
  Phys.\ Rept.\  {\bf 437}, 125 (2007).

\bibitem{Krebs:2008zb}
  H.~Krebs, E.~Epelbaum and U.~G.~Meissner,
  Phys.\ Rev.\  C {\bf 80}, 028201 (2009); 
  arXiv:0905.2744 [hep-th].

\bibitem{Becher:1999he}
  T.~Becher and H.~Leutwyler,
  Eur.\ Phys.\ J.\  C {\bf 9}, 643 (1999).

\bibitem{Gegelia:1999gf}
  J.~Gegelia and G.~Japaridze,
  Phys.\ Rev.\ D {\bf 60}, 114038 (1999);
  J.~Gegelia, G.~Japaridze and X.~Q.~Wang,
J.\ Phys.\ G {\bf 29}, 2303 (2003).

  
  \bibitem{Holstein:1992xr}
  B.~R.~Holstein,
  Comm. Nucl.\ Part.\ Phys.\  {\bf 20}, 301 (1992); See also, B.~R.~Holstein,
  these proceedings.
  
  \bibitem{Federspiel:1991yd}
  F.~J.~Federspiel {\it et al.},
  Phys.\ Rev.\ Lett.\  {\bf 67}, 1511 (1991).

\bibitem{Zieger:1992jq}
  A.~Zieger, R.~Van de Vyver, D.~Christmann, A.~De Graeve, C.~Van den Abeele and B.~Ziegler,
  Phys.\ Lett.\  B {\bf 278}, 34 (1992).

\bibitem{Hal93}
E.~L.~Hallin {\it et al.},
Phys.\ Rev.\ C {\bf 48}, 1497 (1993).

\bibitem{MacG95}
B.~E.~MacGibbon, G.~Garino, M.~A.~Lucas, A.M.~Nathan, G.~Feldman and B.~Dolbilkin,
Phys.\ Rev.\ C {\bf 52}, 2097 (1995).

\bibitem{MAMI01}
V.~Olmos de Leon {\it et al.},
Eur.\ Phys.\ J.\ A {\bf 10}, 207 (2001).

\bibitem{Bab98}
D. Babusci, G. Giordano and G. Matone,
Phys. Rev. C {\bf 57}, 291 (1998).

 \bibitem{PDG2006}
  W.~M.~Yao {\it et al.}  [Particle Data Group],
  J.\ Phys.\ G {\bf 33}, 1 (2006).

  \bibitem{Beane:2004ra}
  S.~R.~Beane, M.~Malheiro, J.~A.~McGovern, D.~R.~Phillips and U.~van Kolck,
  Phys.\ Lett.\  B {\bf 567}, 200 (2003)
  [Erratum-ibid.\  B {\bf 607}, 320 (2005)];
  Nucl.\ Phys.\ A {\bf 747}, 311 (2005).
  
\bibitem{Holstein:2005db}
  B.~R.~Holstein, V.~Pascalutsa and M.~Vanderhaeghen,
  Phys.\ Rev.\  D {\bf 72}, 094014 (2005).



\end{thebibliography}
\end{document}